# Network Structure, Self-Organization and the Growth of International Collaboration in Science

Caroline S. Wagner[*] and Loet Leydesdorff
Amsterdam School of Communications Research (ASCoR)
University of Amsterdam

## Abstract

Using data from co-authorships at the international level in all fields of science in 1990 and 2000, and within six case studies at the sub-field level in 2000, different explanations for the growth of international collaboration in science and technology are explored. We find that few of the explanations within the literature can be supported by a detailed review of the data. To enable further exploration of the role of recognition and rewards as ordering mechanisms within the system, we apply new tools emerging from network science. These enquiries shows that the growth of international co-authorships can be attributed to self-organizing phenomenon based on preferential attachment (searching for recognition and reward) within networks of co-authors. The co-authorship links can be considered as a complex network with sub-dynamics involving features of both competition and cooperation. The analysis suggests that the growth of international collaboration is more likely to emerge from dynamics at the sub-field level operating in all fields of science, albeit under institutional constraints. Implications for the management of global scientific collaborations are explored.

## 1. Introduction

In the literature on the growth of collaboration in science and technology, the influence of the rewards system on collaboration at the international level has not received a great deal of attention. (Whitley 1984) Many studies have demonstrated the growth of international collaboration in science (ICS) by using co-authorships. (Narin 1991; Luukkonen *et al.* 1992 and 1993; Miquel & Okubo 1994; Doré *et al.* 1996; Georghiou 1998; Glänzel 2001). Persson *et al.* (forthcoming) show that citations to articles resulting from international collaborations have grown faster than those referring

---

[*] Correspondence to: C.S. Wagner, ASCoR, Kloveniersburgwal 48 Amsterdam 1012 DX Netherlands, c.s.wagner@uva.nl

to domestic collaborations. Narin *et al.* (1991) has shown that internationally co-authored articles are more highly cited than other articles. Despite this body of evidence, the question of *why* this class of research is growing so quickly or why it has a relatively high impact needs more discussion. A theoretically satisfying explanation of the phenomenon has yet to be offered. (Katz & Hicks 1987; Wagner-Döbler 2001)

This article continues these inquiries along two lines. Drawing upon data (published elsewhere) for 1990 and 2000, different explanations offered within the literature for the increase in international collaboration are discussed. We find that existing explanations are incapable of fully explaining the emerging structure of ICS. In seeking other explanations, we turn to recent literature from the network sciences. Physicists needing large datasets for the study of network dynamics have used scientific co-authorships as a subject. In the process, they have revealed fascinating dynamics about collaborative networks, including the mechanism of preferential attachment as a structuring factor. Using the tools developed by Barabási & Albert (1999) and Newman (2001) we investigate whether these mechanisms also apply at the international level. Our findings reveal the emerging structure of linkages within the international system at the sub-field level, and perhaps move science studies closer towards a theoretically satisfying explanation for the rise of international collaboration.

## 2. A brief review of the literature examining the rise of international collaboration

Over the past 25 years, a number of reasons have been suggested to explain the growth of ICS. One group considers the environment within which science operates: the environmental factors can be divided into internal and external factors. A second group considers the connections within and around science, and these can be divided into those related to the growth of capacity to conduct science (more scientists in more countries are available to cooperate), and those examining the increasing interconnectedness of scientists (within and across countries due to transdisciplinarity and the rise of the information society). Table 1 cross-references these factors, and within the resulting cells, the table shows citations to relevant ideas from the literature.

|  | *Internal to science* | *External to science* |
|---|---|---|
| *Relating to the diffusion of scientific capacity* | Center-periphery theory of lagging countries seeking to cooperate with leading ones (Schott 1998; Shils 1983; Ben-David 1971) | Rising investments by nations and donors increasing S&T capacity (Wagner *et al* 2001) |
| *Relating to the interconnectedness of scientists* | Internal disciplinary differentiation of science (Stichweh 1996); Field-specific characteristics of megascience (Galison & Helvy 1992); Professionalisation of scientific institutes (Beaver & Rosen 1978) | Historical relationships relates to geographic proximity or colonial ties (Zitt *et al*. 2000) Increase in international trade (Ben-David 1971); Growth of information and communications technologies (Gibbons *et al*. 1994; Starr |

Table 1. Factors offered in literature to explain the growth in international collaboration in science

Each of these ideas merit consideration, and they may have been fully explanatory at the time they were offered. Nevertheless, it is argued here that the reasons shown in Table 1 no longer explain the increase in ICS. We have presented data on the shift in network structure at the international level in another paper. (Wagner & Leydesdorff 2004 forthcoming) Key points from this research are discussed below. In light of these findings, it becomes clear that explanations offered in the literature do not sufficiently specify how rewards and incentives (Whitley 1984) serve as a driver of intellectual organization at the global level.

The data created at the global level for 1990 and 2000 show significant shifts in the network structures over the decade. Table 2 inventories the data and shows initial results. (The numbers reflect "integer counting" which attributes a "1" to each occurrence of authorship from a country.)[1] The percent of internationally co-authored articles nearly doubles during the 1990s to account for 15.6 of all articles published. A large core of cooperating countries expands between 1990 and 2000 from 37 to 54 countries. Additional analysis using the cosine allowed us to view relationships that are particularly intense. We found, for example, some country pairs (usually geographically proximate) have a very close relationship in science (i.e., Zaire and Congo, Iraq and Libya). These geographical groupings may be otherwise isolated.

| Year | Unique documents in SCI | Addresses in the file | Authors for all records | Internationally co-authored records | Addresses, internationally co-authored records | Percent internationally co-authored documents |
|---|---|---|---|---|---|---|
| 2000 | 778,446 | 1,432,401 | 3,060,436 | 121,432 | 398,503 | 15.6 |
| 1990 | 590,841 | 908,783 | 1,866,821 | 51,596 | 147,411 | 8.7 |

Table 2. Data on international network of co-authorships, 1990 and 2000
Original data sourced from ISI

Between 1990 and 2000, we found that the global network has expanded (more players are involved), and we show that it has become more interconnected (more links occur between players). The cluster created by scientifically advanced countries has expanded, and new entrants have joined regional networks, but some nations (e.g., the Arab countries in the Middle East) are grouped into otherwise disconnected networks. The analysis further suggests that the network is becoming more decentralized, with regional "hubs" emerging (e.g., South Africa), with a strong core group of collaborating countries growing from 6 to 8 countries.

The increased volume of internationally co-authored publications seems to have reinforced emerging structures at the global level. The global level can be considered as providing increasingly a system of reference other than the national systems. This system is highly structured: A factor analysis reveals that some of the leading countries compete for co-authorship relations with less developed countries. As countries become more scientifically advanced, they become more able to compete for collaborators from smaller or more peripheral countries. As smaller or "peripheral" countries gain scientific capacity, they appear to be able to join the global network.

Political influences and special programs can be seen to have some affect on linkages in the observed network. For example, the existence of special development aid programs between Scandinavian countries and Latin American and African countries may be the reason that links appear between these countries at the global level. In addition, one can surmise that the policy of the European Union, one that heavily favours research proposals that include two or more nations in collaboration, is influencing the growth of the network in that region. Overall, political influences continue to operate but at a lower order of influence as the global system emerges.

The data at the international level allow us to revisit and critique the reasons offered in the literature that have sought to explain the rise in ICS, introduced in Table 1. Let us begin with theories that considered factors *internal* to science. Centre-periphery theory seeks to explain the rise in international collaboration. Schott (1998), following Ben-David (1974) and Shils (1983), sees the progression related to a succession of countries that have acted as "centres" for world science. Countries at the periphery (often smaller countries) emulate the organisation, orientation, and excellence of scientific work at the centre. As they emulate and adapt the practices of the core country, the capacities of the peripheral countries grow. This dynamic may have been at work in the past, but the data on international collaboration presented for the decade of the 1990s suggests that the centre-periphery model of international scientific collaboration can be replaced with a model that accounts for various centres that both collaborate among and compete with one another for partners from smaller national systems. A core group of scientifically advanced countries[2] is both competitive and highly related. At the lowest levels of the hierarchy, smaller, more peripheral countries are more likely to link to the international network through regional hubs rather than through an advanced country.

A second theory approaches the rise in international collaboration by suggesting that internal disciplinary differentiation of science is influencing the organization of ICS (bottom left box in Table 1.) This includes Stichweh's (1996) assertion, following Price (1963) that collaboration arises in scientific disciplines as they become more specialized. As this happens, scientists must look further afield to find collaborators with similar interests. Research presented in Wagner & Leydesdorff (2003) examining disciplinary linkages in the field of geophysics and the more specialized subfield of seismology cannot support this assertion. Although the subfield of seismology is more specialized, geophysics remains the more highly internationalised of the two fields. Moreover, Wagner (forthcoming) shows that, within six case studies, growth in ICS occurs across *all* fields of science, not just those that are highly specialized.

A third theory (also bottom left box in Table 1) suggests that international linkages result from the financial demands of some fields of science. (Galison & Hevly eds. 1992) Indeed, previously, cross-border links in science were considered to be the extension of national systems seeking to complement each other's capabilities. (Crawford 1992) These activities often take the form of "big science" or "megascience" projects, those motivated because the cost of a single project is too large for any one nation to afford. This organizing imperative continues to operate within fields such as fusion research or astrophysics. Nevertheless, cost sharing alone cannot explain the rapid rise of international collaboration. International collaboration is growing in all fields of science, not just in megascience. Constructed collaboration alone cannot explain the rapid rise in international science.

Beaver and Rosen (1978) suggest that collaboration grew historically as science became "professionalised"—taking place in dedicated institutions of science. Using an historical, nationally-based approach, they show that "collaboration becomes a mechanism for both *gaining* and *sustaining* access to recognition in the professional community." (1978) They refute other authors who claim that collaboration is historically recent, or that it is principally a response to specialisation. Collaboration is intrinsically advantageous to scientists, they argue, particularly when it occurs between a "master" and an "apprentice." It is difficult to compare this with data at the global level. But, this theory can be helpful in understanding the dynamics at the sub-field level, so it will be discussed again below.

Among factors *external* to science (right side of matrix in Table 1) is one suggesting that increased public support for research and development in many countries has enhanced scientific capacity and increased the pool of potential collaborators. Indeed, one of us has suggested this in another paper. (Wagner *et al.* 2001) Growth in capacity appears to have an influence on the ability of countries to join the international network. At the observed level, developing countries appear to link more frequently with neighbours at the regional level in 2000 compared to 1990. Nevertheless, capacity alone cannot explain the interest of scientists to participate in international collaboration. The motivation of smaller or developing countries to join the international collaborations can be assumed to be strong, but why are scientists from advanced countries increasingly willing to collaborate across international borders? Even if one were to account for the influence on collaboration of the European Union, the role of the United States and Switzerland among the core cooperating countries would still beg the question.

Historical relationships, colonial ties, and geographic proximity are also offered as reasons for the rise in ICS (lower right box of Table 1). Indeed, regional relations have remained strong where they have been based on cultural patterns, such as within the Francophone community, and in collaboration between small neighbouring states like the Netherlands and Belgium, or Slovenia and Croatia. A rapid increase in ICS is overwhelmingly observable among the member states of the European Union, but these nations are at the same time firmly embedded in a core structure that includes the United States and Switzerland. Secondary networks like the one carried by the Soviet Union and its allies have faded away during the 1990s. Clustering retains features related to geographical proximity and historical relationships, but these are no longer the strongest features affecting links among researchers.

Some have suggested that the availability of the Internet is causing a growth in international collaboration (lower right box in Table 1). While the Internet and information technology generally are enabling factors, research shows that nearly all collaborations begin with a face-to-face meeting. (Laudel 2001) Once collaboration is underway, researchers use the Internet to exchange data and text, but the majority of collaborations begin in the richer communication environment provided at conferences or research sites. This suggests the information and communications technology cannot be considered as a driver for the initiation of collaboration, only as a facilitating agent.

To summarize this review of the literature, the interests of nations to gain efficiencies through collaboration cannot explain the rapid rise in international collaboration in science: collaboration is growing in all fields of science, even where national interests may not be furthered by collaboration.[3] Nor is international science driven solely by the needs of smaller or newer "peripheral" countries to gain access to and imitate the centres of science: even the advanced countries are increasing their participation in international collaboration. Similarly, the increase in scientific capacity particularly among developing nations may be contributing to the growth of international collaboration, but capacity cannot explain the very rapid rise of collaboration that includes the most scientifically advanced nations.[4] Historical linkages once influenced collaboration, but they were less influential in the 1990s than in the past. The rise of the Internet is contributing to increased communications, but it is not a causative factor for geographically dispersed links. If these reasons cannot explain the rapid rise in international collaboration, other reasons that have not been fully explored should be investigated.

## 3. Formulating a new approach

Given the inability of the existing explanations to explain the dynamics observed at the global level in the 1990s, we developed a different hypothesis. We assume that ICS is a self-organizing system creating a network of relationships that can be observed at the communications level. We expect that the mechanism operates at the international level as a factor internal to science at the subfield level, and that the growth in international linkages is due to the mechanism of preferential attachment based on reputations and rewards found within scientific collaboration. We expect that scientists link to others to gain visibility, reputation, complementary capabilities, and/or access to resources. We expect this mechanism to be tied more closely to the intellectual and social organization of science than to other factors (historical ties, proximity, core-periphery attachment). To explore this, we developed data on international co-authorships at the sub-field level and analysed them using network analysis to shed light on organisational dynamics within subfields of science.

We further hypothesized that the networks at the international level grow differently, structured by the drivers of collaboration tied to funding and intellectual organization. Thus, sub-fields of science were chosen to represent each of a set of organizational drivers of collaboration at the international level. The sets of drivers we identified are juxtaposed in Figure 1. The vertical represents organizing features related to funding, from highly organized "top down" activities to spontaneous or "bottom up" activities initiated by researchers themselves. The horizontal represents the location of research, from widely distributed to highly centralized. The juxtaposition creates four quadrants that we explore in the case studies as organizing imperatives influencing ICS. (We acknowledge that, within the research and funding communities, these drivers are not exclusive: top down research can still involve self-organizing teams, and bottom-up research projects still respond to funding incentives, but these categories are offered for heuristic purposes.)

- In the highly organized-centralized "megascience" quadrant, scientists collaborate at or around a central research facility or tool (like CERN). Astrophysics was chosen to represent this quadrant.
- In the highly-organized-distributed "coordinated" quadrant, scientists may be compelled by the circumstances of research or funding to cooperate. They may do so to gain access to data or results coming from a single source, or they may be compelled to collaborate by the requirements set out by the institution funding the research. This would be the case with research funded through the European Union, for example, or through the CGIAR system managed through the World Bank. The Human Genome Project might fit into this category: Virology is the case study chosen to represent this quadrant.

The case studies explicitly sought to test whether self-organizing features, or "bottom up" organization can be shown to be more influential of ICS organization than constructed collaboration. As a result, 2 case studies each were chosen for the quadrants represented by the lower half of the bubble in Figure 1:

- Researchers also self-organize "spontaneously" into collaborative teams from the bottom-up. They may work together to share resources (or they may meet while accessing them) that are relatively rare (localized) such as plants within a rain forest, and this would place them in the "resource-dependent" quadrant; geophysics and soil sciences are the cases used to represent this quadrant.[5]

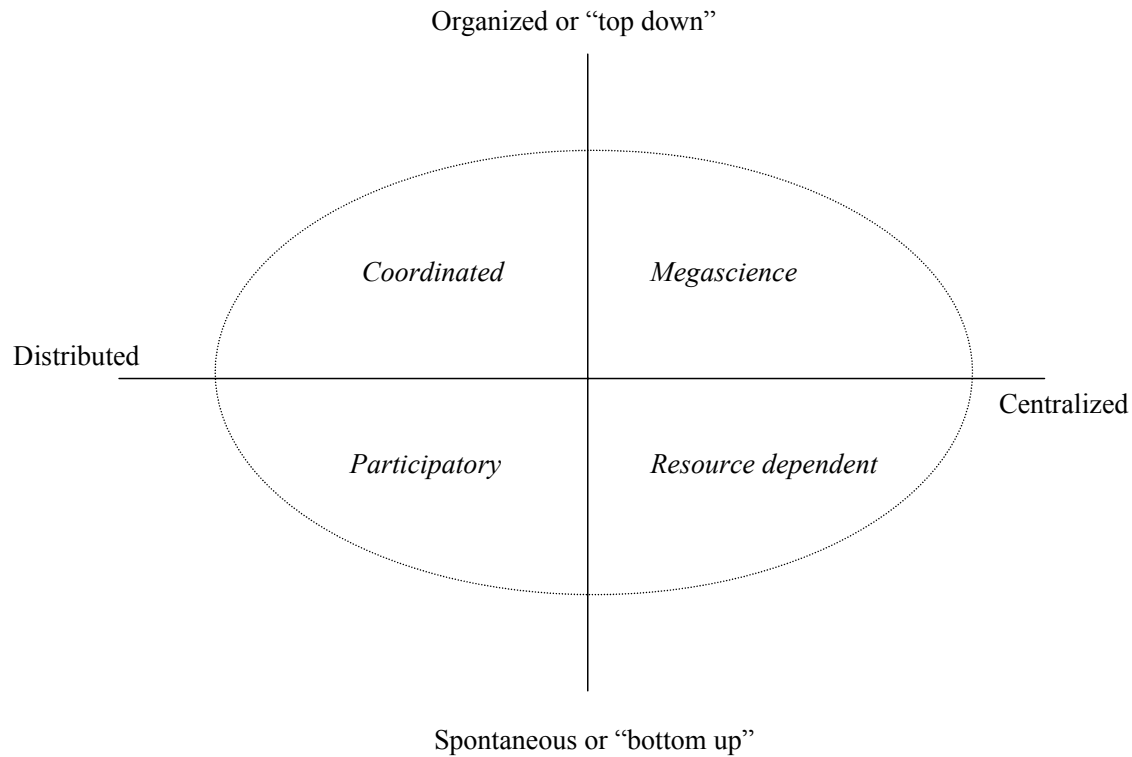

Figure 1. A schematic representation of factors relating to organization of ICS (Source: Wagner *et al.*, 2000.)

- Researchers can self-select collaborators independent of other factors like shared equipment, resources, or the interest of funding institutions, simply because the potential collaborator offers new ideas or complementary capabilities. This type of bottom-up collaboration could be termed "participatory;" the case studies chosen here are mathematics and polymers.

To provide insights that would help us explore these expectations, we created data about ICS, as described in the next section.

## 4. Examining international collaboration at the sub-field level

This paper extends a discussion begun in a separate paper, where we suggest that international collaboration in science is a complex, self-organising network operating according to an internal social dynamic. (Wagner & Leydesdorff 2004 forthcoming) The network is a communications system that operates at one degree of freedom higher than the national level (sub)systems, creating a dynamic that emerges from and then feeds back into lower order subsystems at the national and regional levels. This hypothesis can be explored by examining the growth in ICS at the sub-field level. In order to examine self-organization within the system, it is important to look at the actions of the individual agents within the network. This is done by examining co-authorship data.

Collaborations create a communication system between or among researchers with a shared goal. While it is difficult to observe communication itself, it is possible to measure one outcome of it in the form of shared authorship of a research paper. The addresses of collaborating authors suggest the existence of communications among researchers, and these can be studied as networks of linkages among those who have worked together to create knowledge. (Newman 2001)

## 4.1 Data and methods

Data for six sub-fields of science were drawn from the Institute for Scientific Information (ISI) *Science Citation Index* (SCI) CD-Rom version 2000. The *SCI* is the most reliable source for a comprehensive survey like this one: the country names are standardized.[6] Using a journal set identified through the Journal Citation Reports, and then drawing all related documents from the *SCI* 2000 using Isis's Web of Science, a set of international co-authorships for each field was created.

Clusters of journals related to each other within a citing environment were identified using the *Journal Citation Reports* of the *SCI* (JCR), applying a method detailed in Leydesdorff & Cozzens (1993). For the relevant journals, all articles appearing in the Web of Science for the subject year were downloaded. The author names and addresses appearing in 19,147 articles drawn from 65 journals in six sub-fields were collected. Author names were taken into the database as recorded, no attempt was made to adjust for spelling variants.[7] We did not distinguish among types of contributions (reviews, letters, proceedings, journal articles, etc.) because we are seeking social connections to reveal the structure of the network.

A subset of authors who had co-authored at the international level was created and analysed. The occurrence of authoring events was placed into a pivot table to count the number of times that an author had co-authored papers with a colleague in the sub-field.

These data were drawn into social networks; they were further analysed for a degree distribution among the co-authoring population within the sub-field.

Using the authors as a set of interconnected linkages, the co-occurrence of authorships was fed into Ucinet to examine the degree distribution of authors. For large networks such as these, Ucinet provides the algorithms that find the interrelationships and affiliations within and among networks. (Borgatti *et al.* 2002) The degree distributions were placed into Excel to analyse the power law relationships within the data set. Visualizations of the relationships were created for each sub-field; these are presented below.

### 4.2 Exploring the mechanism of preferential attachment within networks at the subfield level

The emerging field of network science has developed tools to reveal the structural characteristics of highly interconnected systems. (Barabási and Albert 1999; Jeong *et al.* 2001; Newman 2001; Ebel *et al.* 2002; Dorogovtsev and Mendes 2003; Newman 2004) This research has shown that large interconnected networks have features in common, particularly that they display short length-scale clustering (Watts & Strogatz 1998) and that they obey scaling laws. (Albert *et al.* 1999; Barabási & Albert 1999; Jeong *et al.* 2001; Barabási *et al.* 2001) Newman (2000; 2001; 2004) has shown that collaborative scientific networks have a surprisingly short node-to-node distance and a large clustering coefficient (2000), much larger than one would expect from a random network of similar size. In addition, the scientific co-authorship networks that Newman studied have a degree distribution that follows a power law (2001). We applied the tools from network science to examine whether the network of collaborations at the subfield level shares the feature of preferential attachment with other interconnected collaborative networks. (Barabási *et al.* 2002; Newman 2001)

Jeong, Neda and Barabási (2001) find that networks evolve based on two features: growth and preferential attachment. In a study of evolving networks, they show that highly connected nodes increase their connectivity faster than their less connected peers, a phenomenon called preferential attachment. Jeong *et al.* (2001) further show that large evolving networks grow through two mechanisms: the addition of new nodes and new links between existing nodes. As Dorogovtsev and Mendes (2003) say more simply,

"popularity is attractive." In science studies, the observation of preferential attachment has been called the "Matthew effect" recalling the Biblical observation from the Gospel of Matthew that the rich get richer. (Merton 1964)

Building upon this earlier work, Barabási *et al.* (2002) studied scientific collaboration as a complex evolving network. They confirm the findings of Jeong *et al.* (2001) that, over time, the number of nodes in a co-authorship network increase due to the arrival of new authors, and that the total number of links also increases through the connections made between existing authors. They also confirm that node selection is governed by preferential attachment, a feature of scale-free networks in which nodes link with a higher probability to those nodes that already have a larger number of links.

In order to explore whether international collaborative networks operate through the mechanism of preferential attachment across the different subfields of science, our data on co-authorships for six fields of science were examined from this perspective. The six sub-fields represent a range of organizing features, from more fragmented collaboration patterns of mathematics (Dang & Zheng 2003; Grossman & Ion 1995) to the more integrated patterns of physics which has been shown to have highly collaborative network structures (Barabási 2002; Newman 2001; Wagner-Döbler 2001).

We applied a degree-based measure to the data on international collaboration. The degree distribution $P(k)$, giving the probability that a randomly selected node has $k$ links, corresponds to the notion of how well connected an actor is within a network. (Barabási & Albert 1999; Scott 2000; Barabási *et al.* 2002) The degree distribution P($k$) gives the probability that a randomly-chosen actor (node) has $k$ links. A scale-free network is characterized by the following scaling behaviour in $P(k)$:

$$P(k) \sim k^{-\gamma}$$

where γ is the scale exponent. The Barabási-Albert model (1999) shows that many real networks (such as the Worldwide Web) are "scale free," where the value of the exponent ranges from -2 to -3.

The network structure and the co-authorship distribution exposed in the six cases suggest that, at the international level, networks of co-authorship display a preferential attachment mechanism but, in findings similar to Jeong *et al.* (2001), the preference deviates slightly from proportional. Figures 2 through 7 below show the degree distribution for the six cases studied; data is presented in Table 3. Illustrated in these figures is a very high degree of connectedness of authors in the network (the cluster of nodes at the bottom right of the graph.) The figures are shown in a log-log with a power curve as well as a best-fit line. The fit is indicated with the R-square.

In findings similar to others (Barabási & Albert 1999; Albert & Barabási 2000; Barabási *et al* 2002) the networks of collaborations shown in the figures below have fat tailed degree distributions. This has been interpreted as a power-law form. (Barabási & Albert 1999) However, the degree distribution of the international collaborations cannot be fitted into a single power-law dependence. While the exponent falls between 2.3 and 3.6, similar to Barabási *et al.* (2002) and Newman (2001), the power law appears to operate only in the middle of the distribution. The graphs show a hooked end and a fat-tailed distribution. We will offer a possible explanation for these observations below.

| Field of Science | $\gamma$ | $R^2$ | *Cluster Coefficients* | |
|---|---|---|---|---|
| | | | mean | random |
| *Astrophysics* | -2.79 | 0.93 | 0.012528 | 0.000928 |
| *Geophysics* | -2.81 | 0.91 | 0.018750 | 0.000825 |
| *Math logic* | -2.25 | 0.91 | 0.014727 | 0.008109 |
| *Polymer Science* | -3.12 | 0.94 | 0.010356 | 0.000451 |
| *Soil Science* | -2.99 | 0.88 | 0.021444 | 0.001098 |
| *Virology* | -3.61 | 0.95 | 0.017871 | 0.000622 |

Table 3. Degree distribution and cluster coefficients for six ICS case studies.

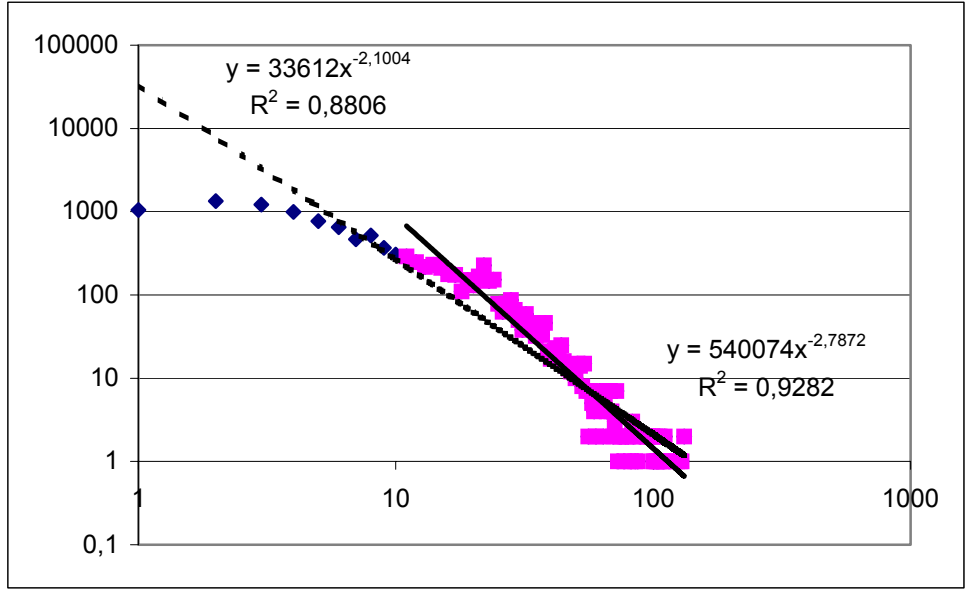

Figure 2. International co-authorship degree distribution in Astrophysics (2000)

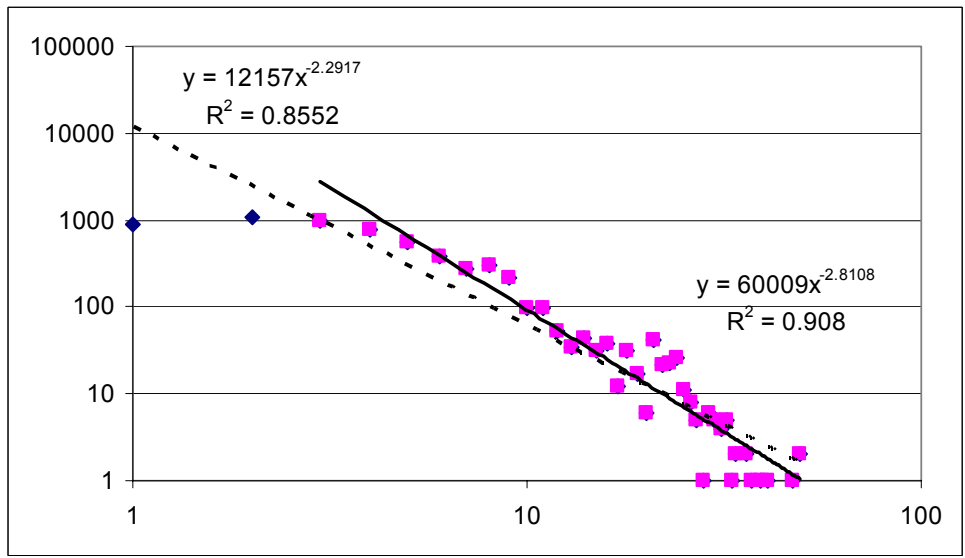

Figure 3. International co-authorship degree distribution in Geophysics (2000)

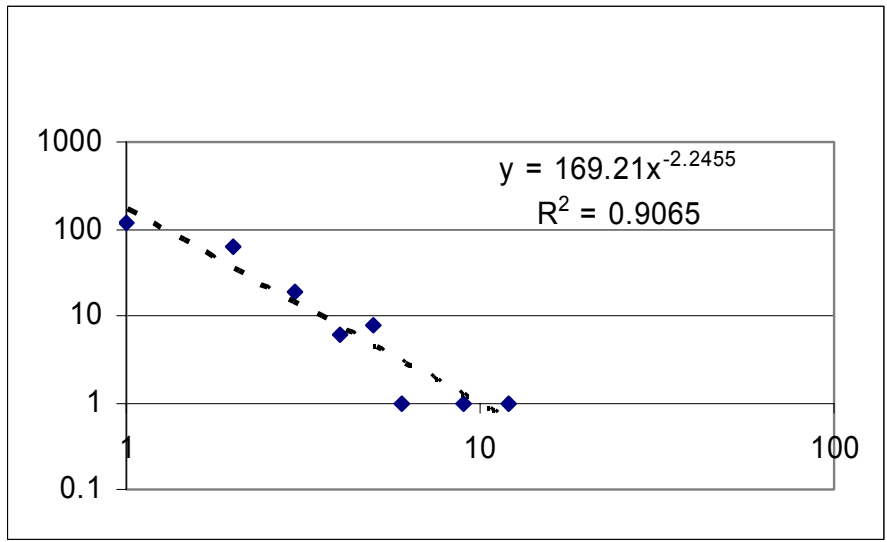

Figure 4. International co-authorship degree distribution in Mathematical Logic (2000)

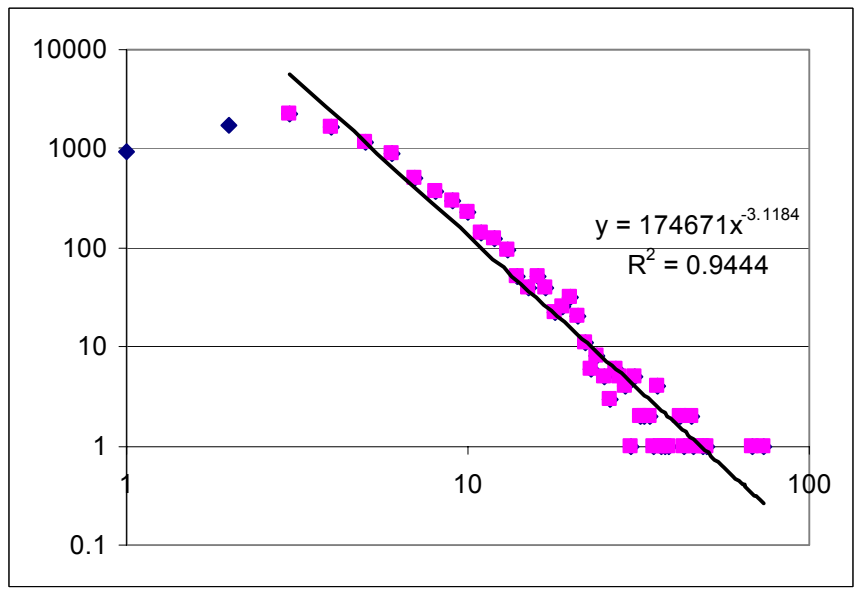

Figure 5. International co-authorship degree distribution in Polymer Science (2000)

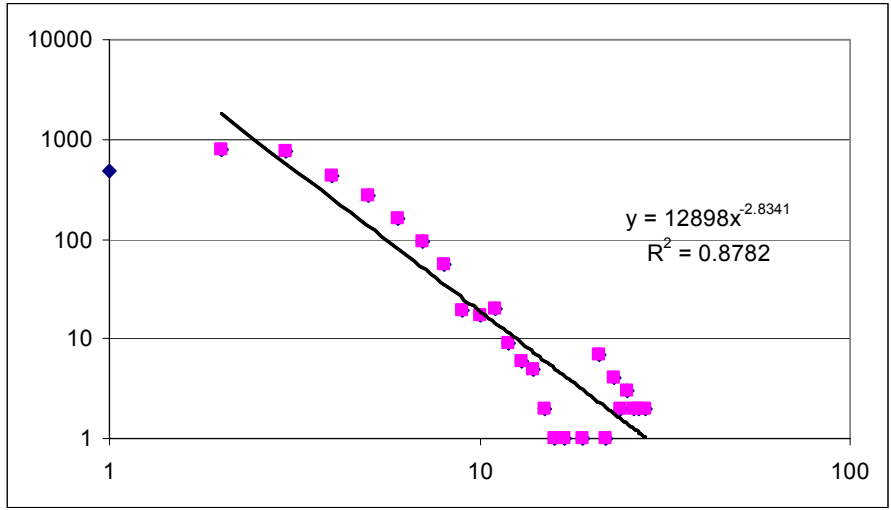

Figure 6. International co-authorship degree distribution in Soil Science (2000)

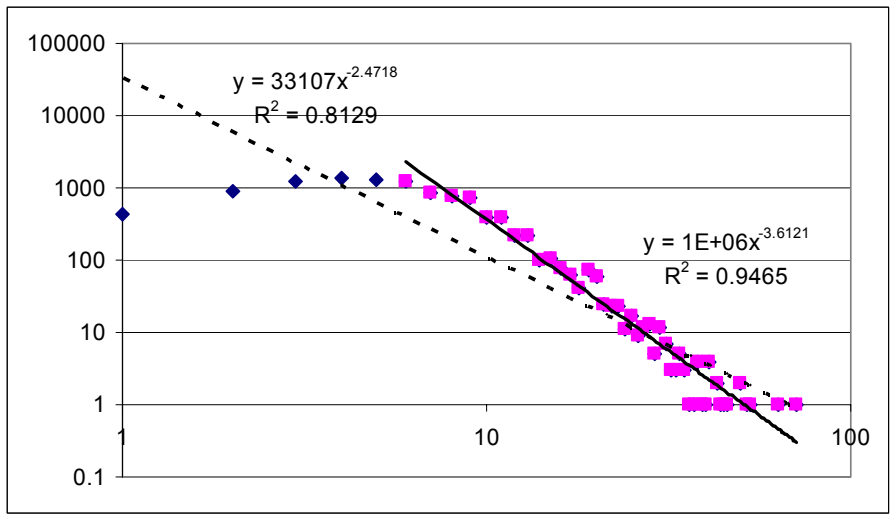

Figure 7. International co-authorship degree distribution in Virology (2000)

These data suggest that the scale-free distribution of co-authorships that Barabási *et al.* (2002) and Newman (2001) found for fields of science in general also holds at the international level. The data further suggest that these networks have "small world" properties. Small worlds are a type of highly-clustered network with short path lengths and high clustering coefficients (See Table 3). (Watts 1999) With the exception of 'mathematical logic,' Table 3 shows that the mean coefficient is orders of magnitude higher than what would be expected in a random network. Small worlds are highly efficient at localised exchanges of information, where they share common information and methods among those operating in the cliques. (Newman 2001; Cowan & Jonard 2003)

With the exception of 'mathematical logic'—which may have been too small a sample (N = 147)—all distributions exhibit a hook and a tail in addition to the power-law configuration. The hook has been noted more often in the literature and various proposals have been made for making the power-law distributions more complex than linear. However, this does not explain the deviation. We submit that both the hook and the tail of the distributions can be considered as the institutional constraints on the dynamics of preferential attachment that prevail in the middle part of the distribution and consequently follow a power-law distribution. The hook can be identified with the arrival of newcomers into the field. This is a condition for the institutional reproduction. The tail can be considered as representing an elite group of scientists who no longer compete uphill for co-authorship relations, but who function as co-authors for the "continuants" given their already established positions.

## 5. The role of "continuants" within the network

To explore the question of how actors select others to work with within the network of possible co-authors, we extend the work of Braun, Glänzel, and Schubert (2001), following Price and Gürsey (1976) on the role of various actors within co-authorship networks. Braun *et al.* built upon the work of Price and Gürsey to explore the dynamics of links within co-authoring communities. Although they were not seeking to show preferential attachment, the data created in these articles is consistent with our findings.

Price and Gürsey provided a scheme for what they called the "actuarial statistics of the scientific community" by categorizing authors to represent their published contribution over time. They determined four categories that represent the co-authoring patterns: "continuants," "transients," newcomers," and "terminators."[8] Looking at data about authorship and co-authorship over time, and most importantly for this argument, is the role of continuants -- those authors whose work is published in the years before and after the year of interest.

Braun *et al.* (2001) found that, for the field of neurosciences, only *continuants* published more than 10 papers in a year. Indeed, these authors find that the preference structure of authors for cooperating with each other is highly skewed: continuants rarely appear as single authors and the overwhelming number of papers is co-authored by continuants. They report that "coauthorship relations among these three categories [newcomers, transients, and terminators] are usually also mediated by continuants." (p. 508)

The co-authorship data and preferential attachment analysis of the six case studies supports and refines Braun *et al.*'s argument. In network terms, their findings could be reinterpreted to say that continuants play a role within the network as "nodes" to which others connect. A large number of continuants are competing for reputations and reward in terms of international coauthorship relations using the mechanism of preferential attachment. However, some become so well connected that network theorists would call them "hubs"—nodes to which lots of people within the network connect or seek to connect. They appear within the power curve at a high degree in a scale-free network (on the bottom right of the figures). In other words, these continuants can be considered as hubs within active small world networks. The hubs no longer compete among themselves upward in terms of adding coauthorship reputation, but compete in terms of building networks of intellectual followers of the next generation.

What role do these hubs play at the field level within scientific communities? Let's assume that the top of the scale of each of the subfields (typified in Figure 2) is occupied by the continuants, as Braun *et al.* found for neurosciences. Figure 8 points to the position of continuants on the scale. We can ask: What is it that these continuants are mediating? Why do they act as attractors for others who are seeking to collaborate? How do they select whom to collaborate with?

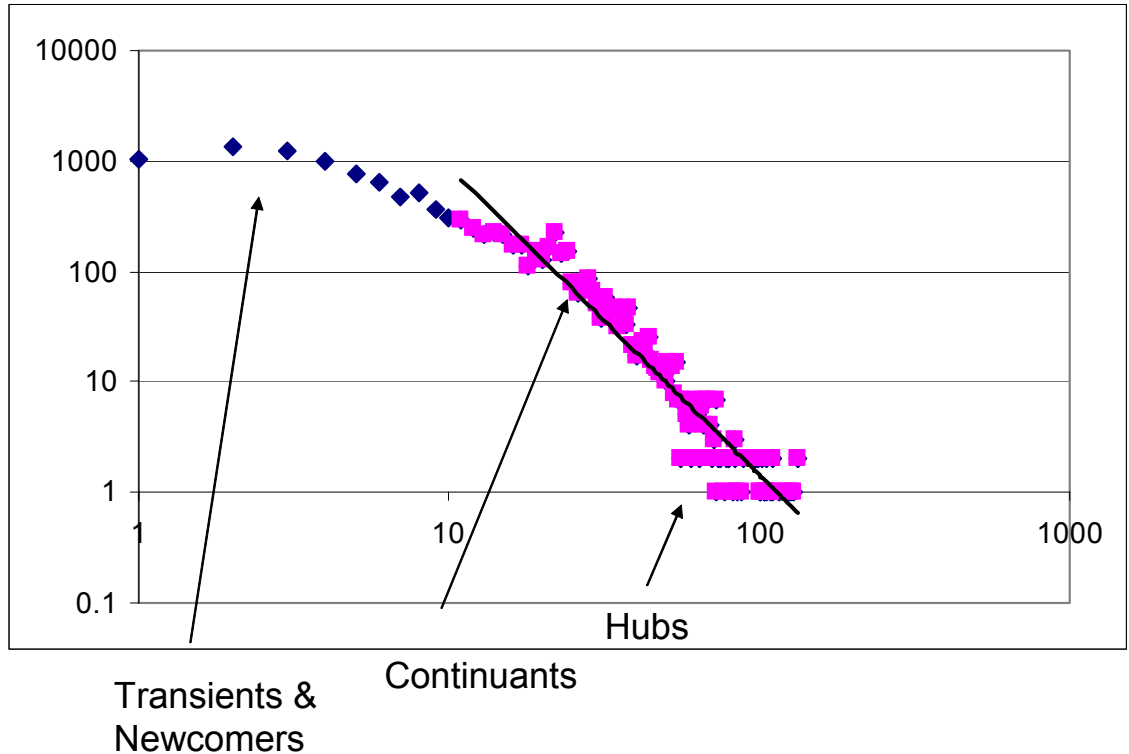

Figure 8. The place of continuants, hubs, and transients and newcomers on the power curve of international co-authorships.

Network theorists suggest that links within small world networks are non-random—there are specific reasons why links are made. Hubs dominate the structure of the networks in which they are present, and they have a specific role to play in networks. They Recall that network theory has found that within networks, actors display preferential attachment: when choosing between two possible links, agents will seek to connect to the more connected node.[9] In other words, when someone is seeking a collaborator, they will seek someone who is already highly connected. This also supports Braun *et al.* in their findings that continuants mediate relationships among co-authors within a field.

This concept of preferential attachment is similar to the findings of Jacob Meyerowitz (1985) who presented a concept of membership within mediated groups, noting the following:

> For a team to have its own sense of identify, its physical location matters less than its "shared but secret information." Members have access to this privileged information. Such information separates members ("us") from others ("them") who do not have the same access….New members become part of a group through "controlled access to group information," the format and information processes of socialization.

The more senior (or published) members of a group hold privileged social and technical information. Newcomers and transients seek access to this information and recognition within their field. They also seek the reflected glory of working with a highly-cited and respected researcher. As a result, the continuants, acting as hubs within their scientific networks, can be said to be attractive as a potential collaborator. They, in turn, can choose carefully among the many opportunities to work with others: As Melin (2000) found, senior researchers would work with junior people to gain higher productivity and credibility within their field. The costs of collaboration are borne by the newcomers and transients who potentially gain greater visibility by working with a well-known person. (Bala & Goyle 2000) The ties formed within fields are mediated by continuants who act as gatekeepers to newer entrants into the network, creating dynamics at the sub-field level.

## 6. Theoretical implications of this research

This article argues that international co-authorships within fields of science can be shown to self-organize based on rules of preferential attachment among productive researchers. The networks examined for this study have self-organizing features, suggesting that the spectacular growth in international collaborations may be due more to the dynamics at the sub-field level created by individual scientists linking together for enhanced recognition and rewards than to other structural or policy-related factors. The choices of individual scientists to collaborate may be said to be motivated by the reward structures of science where co-authorships, citations and other forms of professional recognition lead to additional work and funding in a virtuous circle. Highly visible and productive researchers within the field, able to choose among potential collaborators, choose those most likely to enhance their productivity and credibility. These "continuants" thus mediate the entrance into this network. This creates a competition within a field of science for collaborators.

The competitiveness of this system means that researchers seek to work with those who offer the greatest chance of creating new knowledge. New knowledge creation often results from unstable networks with weak ties, where those within the team are exposed to new ideas and new methods. (Granovetter 1973) The weak ties are relatively easy to make and sever at the international level, where the pool of potential collaborators, already selected by the system, is large enough to offer diversity. The small world nature of the network means that it is relatively easy for these people to know each other's reputations and seek each other out, when needed, for collaborative research.

This framework for thinking about international collaboration supports the work of some researchers, including Braun, Glänzel, Schubert, Barabási, Melin, and the earlier work of Beaver. It suggests that some other explanations discussed here for the rapid growth of international collaboration cannot be supported, notably the centre-periphery theory, the concept that the Internet or trans-disciplinarity are causing the rapid growth, or that cost-sharing is a decisive factor. This theory does not disprove the idea that specialization in science is driving international collaboration, but it also does not support it. The same can be said for the idea that increased capacity is a causative factor for international collaboration: this cannot be shown by the data, but it is also not inconsistent with it.

This theory needs further testing by conducting comparative analysis on international co-authorship at the field level for different fields than the ones shown here, and by modelling this framework and determining if an agent-based model with the basic rules of preferential attachment can create similar outcomes to what is seen in the data. Deeper understanding of the role of enhanced S&T capacity on the ability of developing countries to join ICS could also provide useful insights in whether dynamics differ based on capacity.

## 7. Policy Implications of this Research

We expected to find that the organizational drivers shown in Figure 1 explain the structure and growth of collaboration at the international level. While these organizational drivers continue to operate, this research shows that, during the 1990s, a network structure created by the mechanism of preferential attachment, and operating in all the fields of science examined, has overtaken these drivers to become the most influential factor influencing ICS organization. This finding has significant implications for research policy.

Policymakers take an active interest in ICS from both a political and an epistemic perspective. The political perspective includes using ICS to meet a range of goals, including cost savings, security, capacity building in developing countries, and political goodwill. (Crawford 1992; Skolnikoff 1994; Wagner 1997) Salomon has suggested that ICS can contribute to regional political stability. (2001) Recent reports on international collaboration from governmental and non-governmental organizations (cf. Interacademy Council Report; World Bank reports; United Nations reports) highlight the importance of ICS to encourage growth in developing countries. Epistemic interests are expressed by policies that support large-scale research activities and those that encourage researchers to travel to conferences and take part in international projects. Lack of constraints on public money with regard to international consultation is a passive indicator to researchers to seek cooperation where it is most useful to research and knowledge creation.

Non-government policy-related groups also take an interest in ICS. Development aid institutions like Sweden's Sida, the World Bank, and NATO all fund ICS, each seeking to address specific goals that can include poverty alleviation or other specific problems of the poor in Sida's case, enhanced agricultural productivity in the case of the World Bank, or enhanced security in NATO's case. Similarly, and on a larger scale, the European Commission has centered its research programme on the goal of encouraging intra-European research networks. (Verspagen 2001; Breschi & Cusmano 2003)

Each of these policy mechanisms, tools, and goals continues to operate and influence the formation of ICS. But, to the extent that policies for ICS mimic national policies or use older models for structuring participation in ICST, this research suggests that they may not fully exploit the benefits and opportunities of research taking place at the global level, or worse may find that the system fails to produce the desired outcomes.

The mechanism of preferential attachment is influencing ICS initiation, management, and governance in ways that perhaps are not explicitly considered or well understood now.  This research shows that ICS is a system of communications that requires a research policy approach that would diverge considerably from an institutionally-based or geographically-tied model.  For example, university-based researchers may be working closely with international colleagues but maintaining minimal interaction with juniors in their own institutions.  Similarly, a researcher in one country may find that her results are of interest to a corporate research centre in another country, and work to exploit the results there.  Knowledge is highly portable and researchers are seeking the reward of recognition: the network may change the physical location where knowledge is created, and where it is exploited.  Enhancing policy tools that increase the attractiveness of local researchers and ensure capture of spill-overs may be needed.

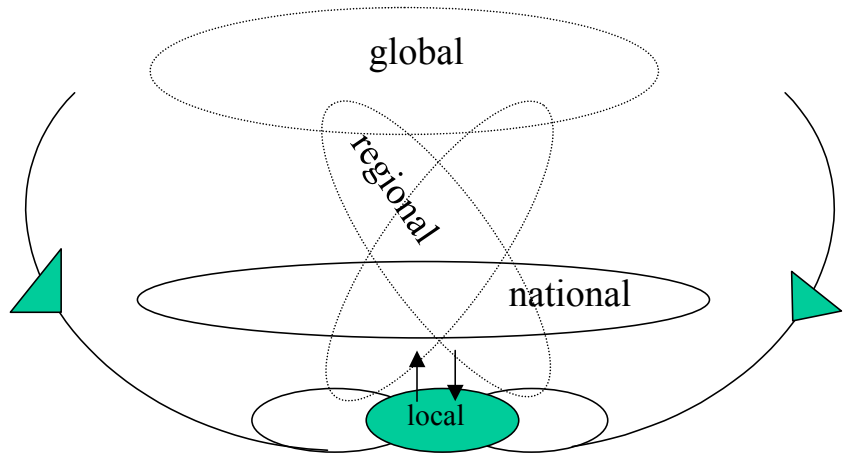

Figure 9. A schematic view of the communications systems operating at different levels

The knowledge-creating community operates on a number of levels, as suggested in Figure 9. The global level exists only as a networked system, but one that feeds back into the national, regional, and local levels, influencing the organization of research at the lower levels. Once linkages are made and research is underway, the knowledge created and shared becomes a catalyst for other links. At this point, ensuring that knowledge can flow freely within the research system is critical to both its growth and expansion, and to the ability of any actor to ensure that knowledge is available to users at the local level. The dynamic shifts from the focus on the *nation* as the system of innovation to a local-global nexus where research carried out locally becomes available within a global system. The national policy structure then becomes either an enabler or an inhibitor of knowledge flows and linkages among actors.

The questions of initiation of research and participation in ICS assume a direct and active role for policymakers. For those areas where government has direct influence, research managers and policymakers may wish to consider how to ensure that researchers are both networking globally as well as locally or regionally. Research conducted at a remote locations can still be useful at the local level, but a knowledge bridge needs to be created that will link to remote science: often this takes place because people move from one place to another. (Cowan & Jonard 2003) Organizations such as the European Commission that have made this a focus of policy may become a model for integrating diverse knowledge systems.

While there remains a direct role for research managers in facilitating large-scale research, this article suggests that much ICS is only influenced by government policy only at the level of incentives. Thus, it is may be important to differentiate between those types of ICS where government has a direct role in building links through funding and institution building (such as megascience or coordinated research activities) and where government has only an indirect role (such as participatory research) by creating attractive researchers or research opportunities. It may be most efficient to create opportunities for researchers to meet around interesting questions in a way that allows them to seek the most efficient and effective knowledge connections and organization. In cases where government has only an indirect role, yet sees advantage from an ICS strategy, incentives for scientists to participate that consider the rewards structure within the sub-field may be important.

This research suggests that the ability of any actor to join the ICS network depends on their *attractiveness* as a partner. A shift from funding research only for those researchers within a prescribed national, regional or other programmatic areas towards funding the best internationally connected research, no matter where it will take place, may actually enhance knowledge creation. While it may be possible to construct collaboration as a policy initiative, it will only be sustainable if is supported by good science and solid technical skills on the ground. For this reason, particularly for developing countries, ICS policy needs to be coordinated with domestic efforts to increase research and development spending and build capabilities. Links should be strengthened among government and research institutions, and individual researchers should be made "stakeholders" in the process of decision-making about ICS investments. Complementary policies to chose targets and tie down locally the knowledge created at the global level may need to become more explicit as ICS continues to grow.

## Acknowledgements

Special thanks to Andrea Scharnhorst, Nederlands Instituut voor Wetenschappelijke Informatiediensten for help with the preferential attachment measures, and to
 Jonathan Cave, University of Warwick, Stuart Blume, University of Amsterdam, and Donald D. Beaver, Williams College, for comments on earlier drafts.

Version 24 June 04

Endnotes

---

[1] Fractional counting attributes the numbers proportionally, so that the number of authors on any given paper reduces the share of each participating country. A second way of counting is to identify the number of *links* represented among the countries involved, with each bilateral relationship counting as "1." This normalization in terms of number of links is more common in network analysis. A third way of counting is integer or whole/distinct count that attributes a count of "1" to each occurrence of authorship by a country created by the participation of researchers from that country.

[2] The concept of scientific capacity, and a parsing of countries of the world by categories such as scientifically advanced, proficient, developing, and lagging, are offered in Wagner *et al.* 2001a.

[3] Collaboration could be seen as reducing the possibility that knowledge spillovers will be exploited within the nation making the investment. For example, many governments list polymers and other new materials as key technologies with the expectation that these sciences will be closely tied to economic growth. Thus, governments have an implicit expectation that investment in this research will provide a national economic advantage and would not like to see any competitive advantages to leak away through international collaboration.

[4] Among the scientifically advanced countries, growth can be seen in collaboration among all the advanced countries, including but not limited to the countries of the European Union, where collaboration has been incentivized through public policy programmes.

[5] For the two "spontaneous" quadrants, we conducted 2 case studies. This is because we were interested in seeing if self-organizing features had significant influence on ICS growth.

[6] The United Kingdom is considered here in its component parts because the *Science Citation Index* is organized in this way. Addresses are provided as England, Scotland, Wales, and Northern Ireland, and each is accordingly handled as a separate unit for the purposes of this analysis.

[7] Newman (2000) found that the error introduced by authors having the same initials and surnames is on the order of a few percent. Some over-counting can also occur because a single author lists more than one address. We estimated the effect of this phenomenon on the data shown here: In 2000 the data created showed 121,432 internationally co-authored documents. Of these, 6,408 (appr. 5%) had more addresses than authors: 20,449 authors and 29,987 addresses.

[8] Transients were defined by Price and Gürsey (1976) as authors publishing in a given year but not before or after. Newcomers are authors publishing in and after the given year but not before, and terminators were authors publishing before and in the given year but not after.

[9] In network terms, this could be defined by saying that a vertex (or hub) acquires new edges (links with other nodes) with a rate proportional to its degree. (Holmes 2003)